\documentclass[sigconf]{acmart}

\usepackage{booktabs} 
\usepackage{subcaption}
\usepackage{balance}
\usepackage{bbm}
\usepackage{amsmath}
\usepackage{amsfonts}
\setcopyright{acmcopyright}
\usepackage{etoolbox}
\usepackage{mathrsfs}
\usepackage{mathtools}
\usepackage{multirow}
\AtBeginDocument{%
	\providecommand\BibTeX{{%
			\normalfont B\kern-0.5em{\scshape i\kern-0.25em b}\kern-0.8em\TeX}}}

\copyrightyear{2020} 
\acmYear{2020} 
\setcopyright{acmcopyright}\acmConference[SIGIR '20]{Proceedings of the 43rd International ACM SIGIR Conference on Research and Development in Information Retrieval}{July 25--30, 2020}{Virtual Event, China}
\acmBooktitle{Proceedings of the 43rd International ACM SIGIR Conference on Research and Development in Information Retrieval (SIGIR '20), July 25--30, 2020, Virtual Event, China}
\acmPrice{15.00}
\acmDOI{10.1145/3397271.3401192}
\acmISBN{978-1-4503-8016-4/20/07}

\begin{document}
	\fancyhead{}
	\balance
	\title{A Transformer-based Embedding Model for Personalized Product Search}
	
	\author{Keping Bi}
	\affiliation{%
		\institution{University of Massachusetts Amherst}
	}
	\email{kbi@cs.umass.edu}
	
	\author{Qingyao Ai}
	\affiliation{%
		\institution{University of Utah}
	}
	\email{aiqy@cs.utah.edu}
	
	\author{W. Bruce Croft}
	\affiliation{%
		\institution{University of Massachusetts Amherst}
	}
	\email{croft@cs.umass.edu}


	\begin{abstract}
		Product search is an important way for people to browse and purchase items on E-commerce platforms.  
		While customers tend to make choices based on their personal tastes and preferences, analysis of commercial product search logs has shown that personalization does not always improve product search quality. Most existing product search techniques, however, conduct undifferentiated personalization across search sessions.
		They either use a fixed coefficient to control the influence of personalization or let personalization take effect all the time with an attention mechanism. The only notable exception is the recently proposed zero-attention model (ZAM) that can adaptively adjust the effect of personalization by allowing the query to attend to a zero vector. Nonetheless, in ZAM, personalization can act at most as equally important as the query and the representations of items are static across the collection regardless of the items co-occurring in the user's historical purchases. Aware of these limitations, we propose a transformer-based embedding model (TEM) for personalized product search, which could dynamically control the influence of personalization by encoding the sequence of query and user's purchase history with a transformer architecture. Personalization could have a dominant impact when necessary and interactions between items can be taken into consideration when computing attention weights. Experimental results show that TEM outperforms state-of-the-art personalization product retrieval models significantly. 
	\end{abstract}

	
	\maketitle

\section{Introduction}
\label{sec:introduction}
Product search systems have been playing an important role in serving customers shopping on online e-commerce platforms in their daily life. Usually, people issue queries about their shopping needs on the platform and purchase items from the search results based on their personal tastes and preferences. 
Aware of this point, 
recent studies have explored to incorporate personalization in product search and achieved compelling results \cite{ai2017learning, guo2019attentive,ai2019zero}. 

Despite its great potential, personalization does not always improve the quality of product search. Based on the analysis of commercial search logs, \citet{ai2019zero} have observed that personalized models can outperform non-personalized models only on the queries where the preferences of individuals significantly differ from the group preference. While applying a universal personalization mechanism sometimes could be beneficial by providing more information about user preferences, especially when the query carries limited information, unreliable personal information could also harm the search quality due to data sparsity and the introduction of unnecessary noise. Therefore, it is essential to determine when and how to conduct personalization under various scenarios. 

Most existing personalized product search models, however, do not conduct differential personalization adaptively under different contexts. \citet{ai2017learning} propose to control the influence of personalization by representing the users' purchase intent with a convex combination between the query embedding and user embedding. This method applies undifferentiated personalization to all search sessions since the coefficient of the combination is a fixed number. \citet{guo2019attentive} fuse query and users' long and short-term preferences to indicate the users' specific intention. While the long and short-term preferences are modeled by attending to the users' recent purchases and a global user vector with the query,  the model itself still conducts personalization all the time.
Later, \citet{ai2019zero} proposed a zero attention model (ZAM) which introduces a zero vector that the query can attend to besides users' previous purchases. In contrast to \cite{guo2019attentive}, by allowing the zero vector to have attention weights, the influence of personalization can be controlled. Nonetheless, despite the ability to adaptively personalize a query-user pair, the maximum personalization ZAM can perform is to equally consider the query and the user information, which may be not enough when the user preference dominates the purchase. 

In this paper, we propose a transformer-based embedding model (TEM) that is more flexible where personalization can vary from no to full effect. As we will demonstrate later in Section \ref{sec:transem}, a single-layer TEM is similar to ZAM but with a larger range of controlling personalization. 
A multiple-layer TEM  takes into consideration the interactions between purchased items so that it could learn potentially better dynamic representations of queries and items, which probably lead to better attention weights. 
We also compare and analyze the ability of personalization between our model and the zero-attention model theoretically in Section \ref{subsec:comp_zeroattn}.  
Our experimental results on the Amazon product search dataset \cite{mcauley2015inferring, van2016learning} show that TEM significantly outperforms state-of-the-art baselines. 



\section{Related Work}
\label{sec:related_work}
\textbf{Product Search. }
Earlier work on product search mainly considers products as structured entities and uses facets for the task \cite{vandic2013facet}. Language model based approaches have been studied \cite{duan2013probabilistic} for keyword search. To alleviate word mismatch problems, more recently, \citet{van2016learning} introduce a latent semantic entity model that matches products and queries in the latent semantic space. 
Learning to rank techniques have also been investigated \cite{karmaker2017application}.
In the scope of personalized product search, 
\citet{ai2017learning} use a convex combination between query and user embeddings for personalization; \citet{guo2019attentive} represent users' long and short-term preferences with an attention mechanism; \citet{ai2019zero} provide insight on when personalization could be beneficial and propose a zero-attention model to control how personalization takes effect. Personalization has also been studied in multi-page product search \cite{bi2019study}. 


\textbf{Transformer-based Retrieval Models. }
Studies on retrieval with transformers have been sparse and most of them leverage pretrained contextual language models, i.e., BERT \cite{devlin2018bert}, which is grounded on the transformer architecture. It achieves compelling performance on a wide variety of tasks such as passage ranking \cite{nogueira2019passage} and document retrieval \cite{dai2019deeper}.  

\begin{figure}
	\includegraphics[width=0.35\textwidth]{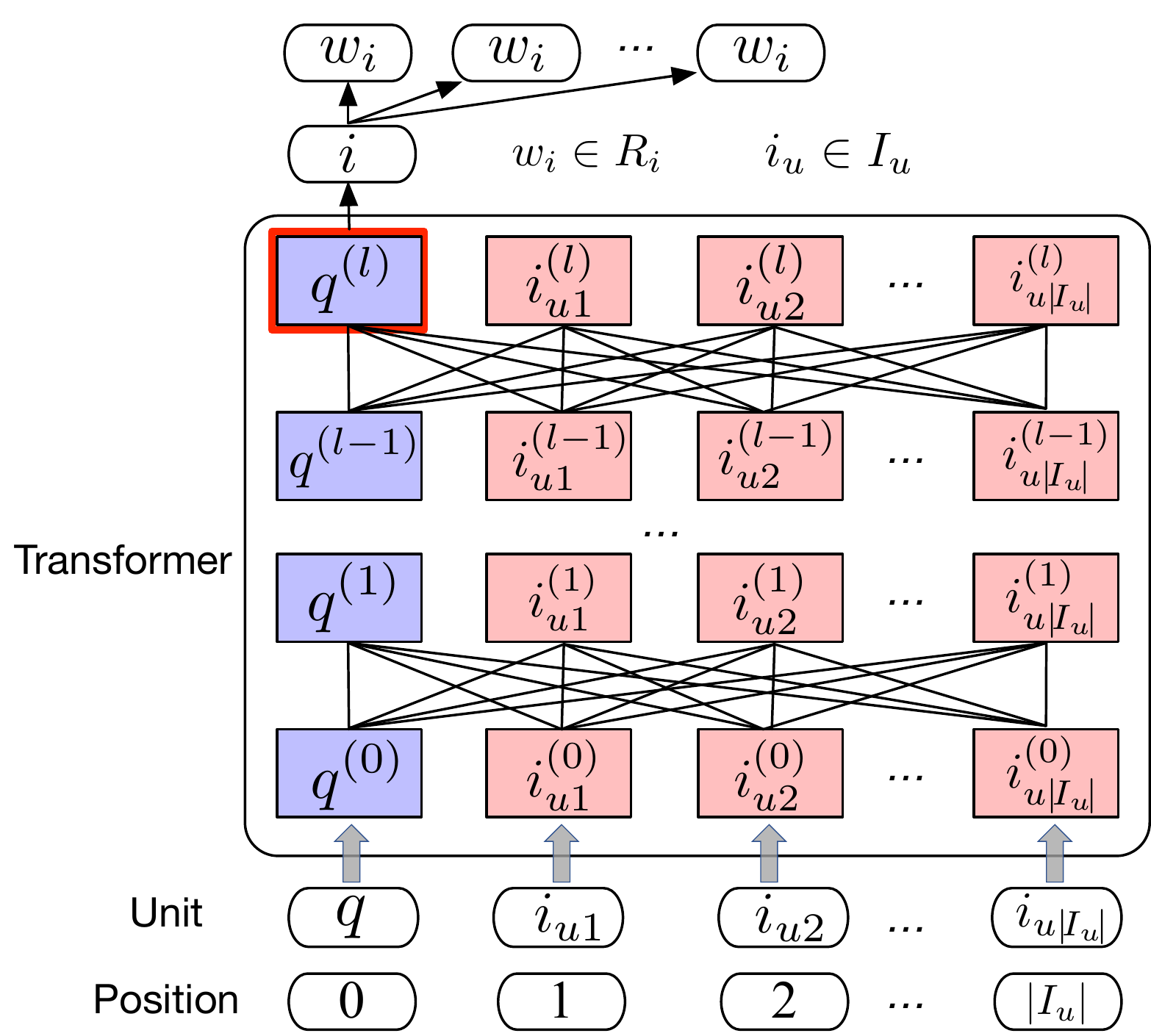} %
	\caption{Our Transformer-based Embedding Model (TEM). }
	\label{fig:tem}
\end{figure}

\section{Transformer-based Embedding Model (TEM)}
\label{sec:transem}
In this section, we first introduce each component of TEM as shown in Figure \ref{fig:tem} and then compare TEM with ZAM theoretically.
\subsection{Item Generation Model}
\label{subsec:itemgen}
We use an item generation model to capture the purchase relationship between an item and query-user pairs. This embedding-based generation framework has been shown to be effective by previous studies on personalized product search \cite{ai2017learning, ai2019zero, bi2019conversational}. 
Formally, let $q$ be a query issued by a user $u$ and $i$ be an item in $S_i$ which is the set of all the items in a collection. The probability of $i$ being purchased by $u$ given $q$ is modeled as 
\begin{equation}
\label{eq:item_gen}
P(i|q,u) = \frac{\exp(\mathbf{i} \cdot \mathbf{M}_{qu})}{\sum_{i' \in S_i} \exp(\mathbf{i}' \cdot \mathbf{M_{qu}})}
\end{equation}
where $\mathbf{i}\in \mathbb{R}^d$ is the vector representation of dimension size $d$ and $\mathbf{M_{qu}}$ is the representation by jointly modeling the query-user pair $(q,u)$. We will elaborate how to yield $\mathbf{M_{qu}}$ later. 

\subsection{Query Representation}
\label{subsec:query_vec}
As shown in previous studies \cite{ai2017learning,ai2019zero,van2016learning},
an effective way to encode query is to apply a non-linear projection $\phi$ on the average query word embeddings:
\begin{equation}
\label{eq:qvec}
\mathbf{q} = \phi(\{w_q| w_q \in q\}) = \tanh(W_{\phi} \cdot \frac{\sum_{w_q\in q}\mathbf{w_q}}{|q|} +b_{\phi})
\end{equation}
where $W_{\phi} \in \mathbb{R}^{d \times d}$ and $b_{\phi} \in \mathbb{R}^{d}$,  $|q|$ is the length of query $q$, and $\mathbf{w_q}\in\mathbb{R}^d$ is the embedding of word $w_q$ in $q$.
This way of encoding queries has outperformed other techniques such as using average word embeddings and applying recurrent neural networks on the word embedding sequence for product search \cite{ai2017learning}. 

\subsection{Item Language Model} 
\label{subsec:item_lang}
As in \cite{ai2017learning, bi2019conversational}, item embeddings are learned from their associated reviews. Let $R_i$ be the set of words in the reviews associated with item $i$. Embeddings of words and items are optimized to maximize the likelihood of observing $R_i$ given $i$: 
\begin{equation}
\label{eq:item_lang}
P(R_i|i) = \prod_{w \in R_i} \frac{\exp(\mathbf{w}\cdot \mathbf{i})}{\sum_{w' \in V}\exp(\mathbf{w'} \cdot \mathbf{i})}
\end{equation}
where $V$ is the vocabulary of words in the corpus.

\subsection{Transformer-based Personalization}
Different queries may need various degrees of personalization \cite{ai2019zero}. Some can be satisfied with popular items in general and some correlates closely with users' historical purchases. 
To represent the purchase intent with query-dependent personalization, we leverage a transformer encoder \cite{vaswani2017attention} architecture to capture the interaction between query and users' historical purchased items, as shown in Figure \ref{fig:tem}. Let $I_u=(i_{u1}, i_{u2}, \cdots, i_{u|I_u|})$ be the sequence of items purchased by $u$ in a chronological order, the size of which is $|I_u|$. We feed the sequence $(q, I_u)$ as the input to a $l$-layer transformer encoder. Since a recent purchase may play a different role compared with  a long-ago purchase, in addition to query and item embeddings of a corresponding unit (query or item), positional embeddings ($PosEmb$) are used to indicate the purchase order of each item. The input vectors to the transformer are:
\begin{equation}
\label{eq:transformer_input}
\mathbf{q}^{(0)} = \mathbf{q} + PosEmb(0);  \text{     }
\mathbf{i_{uk}^{(0)}} = \mathbf{i_{uk}} + PosEmb(k), i_{uk} \in I_u
\end{equation}
where $\mathbf{q}$ and $\mathbf{i_{uk}}$ can be computed according to Eq. \ref{eq:qvec} \& \ref{eq:item_lang} respectively. 

Then user $u$'s purchase intent given $q$, i.e., $M_{qu}$, can be represented with the output vector of query $q$ at the $l$-th layer, i.e., 
\begin{equation}
\label{eq:qu_intent}
\mathbf{M_{qu}} = \mathbf{q^{(l)}}
\end{equation}
We use $\mathbf{q^{(l)}}$ as $M_{qu}$ because it is computed by attending to each transformer input using query $q$ which is more reasonable than other output vectors that attend to the input with a previously purchased item. Specifically, $\mathbf{q^{(l)}}$ is computed as a weighted combination of embeddings of query and purchased items from the previous transformer layer followed by a projection function: 
\begin{equation}
\label{eq:tem_qu}
\begin{aligned}
&\!\!\! \mathbf{q^{(l)}}  \!\!=\! g \Big( \frac{\exp\! \big( \! f\!( q_{Q}^{(\!l-1\!)}, q_{K}^{(\!l-1\!)})\!\big)\!}{\exp\!\big(\!f( q_{Q}^{(\!l-1\!)},  q_{K}^{(\!l-1)\!})\!\big)\! \! + \!\! \sum_{ i' \! \in I_u}\!\! \exp \!\big(\! f\!( q_{Q}^{(\!l-1\!)}, i_{K}^{'(\!l-1\!)})\!\big)}\! \! \cdot \! q_{V}^{(l-1)}  \\
& \;\; + \!\!\! \sum_{ i\in I_u} \!\! \frac{\exp\! \big( \! f\!( q_{Q}^{(\!l-1\!)}, i_{K}^{(\!l-1\!)})\!\big)\!}{\exp\!\big(\!f( q_{Q}^{(\!l-1\!)},  q_{K}^{(\!l-1)\! })\!\big)\! \! + \!\! \sum_{ i' \! \in I_u}\!\! \exp \!\big(\! f\!( q_{Q}^{(\!l-1\!)}, i_{K} ^{'(\!l-1\!)})\!\big)\!} \! \cdot \! i_{V}^{(l-1)} \! \Big) \\
\end{aligned}
\end{equation}
In Eq. \ref{eq:tem_qu}, $f(x,y)$ computes attention score of $y$ with respect to $x$. As in \cite{vaswani2017attention}, $g$ is a projection function that firstly applies $f$ for multiple attention heads followed by a feed-forward layer and residual connections for both the multi-head attention sub-layer and the feed-forward sub-layer. 
$q_{Q}^{(l-1)}, q_{K}^{(l-1)}$, and $q_{V}^{(l-1)}$ are computed according to:
\begin{equation}
q_{Q}^{(l-1)}=\mathbf{q}^{(l-1)}W_q^{Q}; q_{K}^{(l-1)}=\mathbf{q}^{(l-1)}W_q^{K}; q_{V}^{(l-1)}=\mathbf{q}^{(l-1)}W_q^{V}
\end{equation}
where $W_q^{Q} \in \mathbb{R}^{d\times (d/h)}$, $W_q^{K}\in \mathbb{R}^{d\times (d/h)}$ and $W_q^{V}\in \mathbb{R}^{d\times (d/h)}$
are projection matrices; $\mathbf{q^{(l-1)}}$ is the embedding of $q$ at the $(l-1)$-th transformer layer; and $h$ is the number of attention heads. $i_{K}^{(l-1)}$ and $i_{V}^{(l-1)}$ in Eq. \ref{eq:tem_qu} are computed similarly based on $\mathbf{i^{(l-1)}}$, i.e., the vector of $i$ at $(l-1)$-th layer. In this way, TEM can have the capability of ZAM to coordinate personalization and is more general and flexible, as shown in the next section where we will illustrate the relation between TEM with ZAM. 

\subsection{Comparison with Zero-attention Model}
\label{subsec:comp_zeroattn}
In ZAM \cite{ai2019zero}, query and user are jointly modeled by: 
\begin{equation}
\label{eq:zero_attn_qu}
\mathbf{M_{qu}} = \mathbf{q} + \sum_{ i\in I_u} \frac{\exp \big(f'(q,i) \big)}{\exp \big(f'(q,\mathbf{0})\big) +\sum_{ i'\in I_u}  \exp \big(f'(q, i') \big)} \cdot \mathbf{i}
\end{equation}
where $f'$ is a multi-head attention function. 
In ZAM, when $q$ does not require personalization or it has no useful purchase history to attend to, all the items in $I_u$ would have small attention weights, which allows $M_{qu}$ to include information only from $q$. When personalization has great potential for $q$, most attention is allocated to the historical purchases $I_u$ rather than the zero vector. Eq. \ref{eq:zero_attn_qu} shows that the maximum personalization ZAM can conduct is to consider $I_u$ equally important to $q$. However, in some cases, personalization could have a larger impact than queries. \citet{ai2017learning} has shown that the optimal query weight could be much lower than the user weight on some product categories where personalization is indispensable. 
In contrast, TEM based on Eq. \ref{eq:tem_qu} can learn to balance the influence of personalization for each query automatically without limits on the personalization degree. 
Specifically, the query weight can be as small as 0 when personalization is dominant and as large as 1 when personalization is not needed at all. 

In addition, when $l=1$, $\mathbf{q^{(l-1)}}$ and $\mathbf{i^{(l-1)}}$ in Eq. \ref{eq:tem_qu} become $\mathbf{q^{0}}$ and $\mathbf{i^{(0)}}$ (shown in Eq. \ref{eq:transformer_input}) respectively. In this case, the only difference between query and items representations of TEM and ZAM is the positional embeddings. 
When $l>1$, $\mathbf{q^{(l-1)}}$ and $\mathbf{i^{(l-1)}}$  are learned from previous transformer layers by interacting with all the units in the sequence $(q, I_u)$. In this way, the query and items are dynamically represented depending on its interaction with the other units associated with this q-u pair rather than having static vectors across the corpus. By considering the relation between historical purchased items, e.g., same brands or categories, TEM could learn potentially better representation to facilitate product search.





\section{Experiments}
\label{sec:exp}

\textbf{Datasets.}
We use the Amazon product search dataset \cite{mcauley2015inferring} for experiments, as in previous work \cite{ai2017learning, van2016learning, bi2019conversational}. Since there are no available queries for this dataset, we construct queries for each item following the same strategy as in \cite{van2016learning,ai2017learning,bi2019conversational}. A query string of each purchased item is formed by concatenating words in the multi-level category of the item and removing stopwords as well as duplicate words. In this way, there could be multiple queries for each item since an item may belong to multiple categories. The user and each query associated with her purchased item are considered as the possible query-user pairs that lead to purchasing the item. We use three categories of different scales for experiments, which are $Cellphones \& Accessories, Sports \& Outdoors$ and $Movies \&TV$. The statistics are shown in Table \ref{tab:stats}. 

\textbf{Evaluation.}
We randomly divide 70\% of all the available queries into the training set and the rest 30\% queries are shared by validation and test sets. If all the queries of a purchased item fall in the test set, we randomly put one query back to the training set.  We partition the purchases of a user to the training/validation/test set according to the ratio 0.8/0.1/0.1 in a chronological order.  For any purchase in the validation or test set, if none of the queries associated with the purchased item are in the query set for validation and test, this purchase will be moved back to the training set. Our partition ensures that the purchases in the test set happen after the purchases in the training set and no test query has been seen in the training set. 
We use MRR, Precision, and NDCG at 20 as the metrics. 

\begin{table}
	\caption{Statistics of the Amazon datasets.}
	\centering
	\label{tab:stats}  
	\small
	\begin{tabular}{p{0.8cm}  r  r  r }
		\hline
		Dataset & Cell Phones & Sports & Movies \\
		\hline
		\#Users & 27,879 & 35,598 & 123,960 \\
		\#Items & 10,429 & 18,357 & 50,052 \\
		\#Reivews & 194,439 & 296,337 & 1,697,524 \\
		\#Queries & 165 & 1,543 & 248 \\
		\hline
	\end{tabular}
\end{table}

\begin{table*}
	\caption{Comparison between the baselines and our proposed TEM. `*' marks the bast baseline performance. `$\dagger$' indicates significant improvements over all the baselines in paired student t-test with $p<0.05$. }
	\label{tab:overallperf}
	\small
	\begin{tabular}{ c | l || l | l | l || l | l | l || l | l | l   }
		\hline
		\multicolumn{2}{c||}{Dataset}& \multicolumn{3}{c||}{Cell Phones \& Accessories} & \multicolumn{3}{c||}{Sports \& Outdoors } & \multicolumn{3}{c}{Movies \& TV} \\
		\hline
		\multicolumn{2}{c||}{Model} & $MRR$ & $NDCG@20$ & $P@20$ &  $MRR$ & $NDCG@20$ & $P@20$ & $MRR$ & $NDCG@20$ & $P@20$\\
		\hline
		\multirow{2}{*}{Non-personalized} &
		LSE & 0.013 & 0.022    & 0.004  & 0.010 & 0.021 & 0.004 & 0.010 & 0.015 & 0.002 \\
		& QEM &  0.029 & 0.036 & 0.003 & 0.031 & 0.044 & 0.006 & 0.004 & 0.006 & 0.001\\
		\hline
		\multirow{4}{*}{Personalized} &
		HEM & 0.044* & 0.057*    & 0.006* &  0.032 & 0.049 & 0.007* & 0.007 & 0.011 & 0.002\\
		& AEM & 0.043 & 0.049 & 0.004 & 0.031 & 0.045 & 0.006 & 0.013* & 0.020 & 0.003* \\
		& ZAM & 0.041 & 0.046 & 0.004 & 0.040* & 0.057* & 0.007* & 0.013* & 0.022* & 0.003* \\
		\cline{2-11}
		& TEM & \textbf{0.056}$^{\dagger}$ & \textbf{0.072}$^{\dagger}$ & \textbf{0.007}$^{\dagger}$ & \textbf{0.049}$^{\dagger}$  & \textbf{0.074}$^{\dagger}$  & \textbf{0.010}$^{\dagger}$ & \textbf{0.020}$^{\dagger}$ & \textbf{0.028}$^{\dagger}$ & \textbf{0.004}$^{\dagger}$ \\        
		\hline
	\end{tabular}
\end{table*}

\textbf{Baselines.} We include five representative product search models as baselines: the Latent Semantic Entity model (LSE) \cite{van2016learning} which is an embedding-based non-personalized model; Query Embedding Model (QEM) \cite{ai2019zero}, another non-personalized model, which conducts item generation (Sec. \ref{subsec:itemgen} \& \ref{subsec:query_vec}) based on the query alone; Hierarchical Embedding Model (HEM) \cite{ai2017learning} which balances the effect of personalization by applying a convex combination of user and query representation; the Attention-based Embedding Model (AEM) \cite{ai2019zero} which constructs query-dependent user embeddings by attending to users' historical purchases with query, similar to the attention model proposed by \citet{guo2019attentive};  a state-of-the-art model: the Zero Attention Model (ZAM) \cite{ai2019zero} which introduces a zero vector to AEM so that the influence of personalization can be differentiated for various queries. We only include neural models as our baselines since term-based models have been shown to be much less effective for product search in previous studies \cite{ai2017learning, bi2019conversational, ai2019zero}. 

\textbf{Training.} We train our model and all the baselines for 20 epochs with 384 samples in each batch. We set the embedding size of all the models to 128 and sweep the number of attention heads $h$ from \{1,2,4,8\} for attention-based models. The number of transformer layers $l$ is chosen from \{1,2,3\} and the dimension size of the feed-forward sub-layer of the transformer is set from \{96, 128, 256, 512\}. Adam with learning rate 0.0005 is used to optimize the models.

\textbf{Results.}
Table \ref{tab:overallperf} shows the ranking performance of the baseline models and TEM \footnote{The numbers in Table \ref{tab:overallperf} are smaller than those reported by \citet{ai2017learning} since they randomly split user purchases to training and test set which makes the prediction of purchases in their test set easier than predicting future purchases in our test set. }.
Similar to previous studies \cite{ai2017learning, ai2019zero, bi2019conversational}, we observe that LSE and QEM perform worse than personalized product search baselines in most cases.
If we compare the personalized product search baselines, HEM has the best performance on \textit{Cell Phones} whereas ZAM performs the best on \textit{Sports} and \textit{Movies}. Specifically, HEM and AEM achieve better results than ZAM on \textit{Cell Phones} and worse results on the other two datasets. This indicates that, while adjusting the influence of personalization with the attention weights on the zero vector could benefit the retrieval performance of ZAM, its limitation on personalization (i.e., the personalization weight can be no larger than the query weight) could harm the search quality on datasets where personalization is essential.

On all the categories, TEM achieves the best performance in terms of all the three metrics. The improvement upon the best baseline on each dataset is approximately 20\% to 50\%. From the improvement of Precision, NDCG, and MRR, we can infer that TEM not only retrieves more ideal items in the top 20 results but also promotes them to higher positions. This demonstrates that TEM can benefit the effectiveness of personalized models with a more flexible mechanism to control the influence of personalization and by learning dynamic item representations with the interaction between items taken into consideration. 

\textbf{Effect of Layer Numbers}. 
We varied the number of transformer layers to see whether a single-layer or multi-layer transformer will lead to better results on each dataset. The best performance of TEM is achieved when $l$ in Eq. \ref{eq:tem_qu} is set to 2 on \textit{Sports} and 1 on \textit{Cell Phones} as well as \textit{Movies}. 
This indicates that considering the interactions between items does benefit the personalized product search models in some product categories.
\section{Conclusion and Future Work}
\label{sec:conclusion}
In this paper, we propose a transformer-based embedding model, abbreviated as TEM, that can conduct query-dependent personalization. By encoding the sequence of the query and users' purchase history with a transformer architecture, the effect of personalization can vary from none to domination. We theoretically compare TEM with ZAM \cite{ai2019zero} and show that a single-layer TEM is an advanced version of ZAM with more flexibility and a multi-layer TEM extends the model with stronger learning abilities by incorporating the interactions between items co-occurring in users' purchase history. Our experiments empirically demonstrate the effectiveness of TEM by showing that TEM outperforms the state-of-the-art personalized product search baselines significantly. 
For future work, we consider studying TEM for explainable product search. 
The attention scores in TEM indicate the personalization degree and which historical items draw more attention for retrieving a result. This information could be helpful for users to make purchase decisions. In addition, we are also interested in incorporating other information about products such as price, ratings, and images with a transformer architecture to facilitate personalized product search. 

\begin{acks}
This work was supported in part by the Center for Intelligent Information Retrieval. Any opinions, findings and conclusions 
or recommendations expressed in this material are those of the authors and do not necessarily reflect those of the sponsor.
\end{acks}

	\bibliographystyle{ACM-Reference-Format}
	
	\bibliography{reference}
	
\end{document}